\documentclass{article}

\usepackage{spconf,amsmath,graphicx}
\usepackage{footnote}
\makesavenoteenv{tabular}
\makesavenoteenv{table}
\usepackage{caption}
\usepackage{subcaption}
\usepackage{todonotes}
\usepackage{hyperref}
\usepackage{cleveref}
\usepackage{amsmath}
\usepackage{amssymb}

\usepackage[finalnew]{trackchanges}

\usepackage[acronym, shortcuts, nohypertypes={acronym}]{glossaries}
\newacronym{ANN}{ANN}{\textit{artificial neural network}}
\newacronym{Adam}{Adam}{\textit{Adaptive Moment Estimation}}
\newacronym{ASC}{ASC}{\textit{acoustic scene classification}}
\newacronym{CNN}{CNN}{\textit{convolutional neural network}}
\newacronym{OOD}{OOD}{\textit{out-of-domain}}
\newacronym{IID}{ID}{\textit{in-domain}}
\newacronym{SGD}{SGD}{\textit{stochastic gradient descent}}
\newacronym{PCC}{PCC}{\textit{Pearson correlation coefficient}}
\newacronym{KFAC}{KFAC}{\textit{Kronecker-factored approximate curvature}}
\newacronym{PANN}{PANN}{\textit{pre-trained audio neural network}}
\usepackage[activate]{microtype}
\newacronym{GDTUO}{GDTUO}{\textit{gradient descent: the ultimate optimiser}}

\def\x{{\mathbf x}}

\title{Bringing the Discussion of Minima Sharpness to the Audio Domain:\\
A Filter-Normalised Evaluation for Acoustic Scene Classification}
\name{Manuel Milling$^{1,2}$, Andreas Triantafyllopoulos$^{1,2}$, Iosif Tsangko$^2$, \\ \textit{Simon David Noel Rampp}$^2$, \textit{Bj\"orn Wolfgang Schuller}$^{1,2,3}$}
\address{$^1$CHI -- Chair of Health Informatics, MRI, Technical University of Munich, Germany \\
$^2$Chair of Embedded Intelligence for Health Care and Wellbeing, University of Augsburg, Germany \\
  $^2$GLAM -- Group on Language, Audio, \& Music, Imperial College London, UK}
\begin{document}
\ninept
\maketitle
\begin{abstract}
The correlation between the sharpness of loss minima and generalisation in the context of deep neural networks has been subject to discussion for a long time.
Whilst mostly investigated in the context of selected benchmark data sets in the area of computer vision, we explore this aspect for the acoustic scene classification task of the DCASE2020 challenge data.
Our analysis is based on two-dimensional filter-normalised visualisations and a derived sharpness measure.
% with an adjusted application of the \ensuremath{\epsilon}-sharpness.
%After investigating the robustness against the choice of random visualisation directions, 
%We discover that, playing into the dispute in the literature regarding better generalisation of flat minima, in our study, sharper minima tend to show better generalisation than flat minima. 
Our exploratory analysis shows that sharper minima tend to show better generalisation than flat minima --even more so for out-of-domain data, recorded from previously unseen devices--, thus adding to the dispute about better generalisation capabilities of flat minima. 
%This effect is even stronger for the generalisation to .
%We discover that in our study, sharper minima tend to show better generalisation than flat minima, thus adding to the dispute about better generalisation capabilities of flat minima. 
%This effect is even stronger for the generalisation to out-of-domain data, recorded from previously unseen devices.
We further find that, in particular, the choice of optimisers is a main driver of the sharpness of minima and we discuss resulting limitations with respect to comparability. 
Our code, trained model states and loss landscape visualisations are publicly available.
\end{abstract}
\section{Introduction}
When training \acp{ANN} on a specific task, one of the key challenges lies in the network's ability to generalise to unseen data.
As can be interpreted from the universal approximation theorem \cite{hornik1989multilayer}, \acp{ANN} are well capable of representing the underlying data distribution of any task.
In practice --especially given a network with enough depth-- good fits of the training data with converging loss values and perfect evaluation metrics are often easy to find.
However, this does not translate to unseen data, as the generalisation error can vary hugely for almost perfect training loss and can be influenced by the amount of training data, the choice of network architecture, optimiser or batch size \cite{keskar2017on}, among other things.
Models with a high generalisation gap are considered to be overfitted and often perform even worse if the unseen data is \ac{OOD}.
%This effect of overfitting often increases even more if the unseen data is \ac{OOD}.
% , which is not available at training time.
This can, for instance, be observed in the yearly DCASE \ac{ASC} challenge, in which the organisers added new recording conditions, such as different recording devices or cities, only to the test data.
%This can for instance be observed in the yearly DCASE  challenge, in which the organisers added different recording conditions, such as new recording devices or cities only in the test data.
% For the selection of models it is thus common to put emphasis on the performance of model checkpoints on the validation data during the training, thus neglecting any information present in the loss landscape of the training.
%AT:
Critically, model selection, in the form of choosing hyperparameters or `early stopping', is predominantly performed based on validation performance, which on its own can bring quite some limitations as, for instance, reported for \ac{OOD} performance \cite{Triantafyllopoulos2021Fairness}. 
%which often comes from the same distribution as the test data.
%This fails to account for \ac{OOD} samples.
%Yet, this is an important omission, given the fact that in-domain performance is not necessarily indicative of \ac{OOD} performance\todo{underspecification}.

An alternative perspective on model states can be gained by examining the behaviour of loss functions.
Specifically, some characteristics of a model state's minimum have been pointed out to show an important connection to the generalisation error.
% Nevertheless, a more in-depth look into the behavior of loss functions 
% AT
%This has shown some important connections between the characteristics of minima, to which models converge, and the corresponding generalisation capabilities. 
%The terms 
% AT:
\textit{Flatness} and \textit{sharpness} play a particular role here, with flatter minima often believed to have better generalisation \cite{cha2021swad}, at least since the work of Hochreiter and Schmidhuber \cite{hochreiter1997flat}.
Intuitively, these terms are related to the Hessian matrix, which contains all second-order derivatives, at a given point of a function, for all directions and can thus represent the local curvature behaviour of the function.
Yet, an undisputed definition %allowing for efficient implementations 
of flatness and sharpness in the high-dimensional parameter space of \acp{ANN} is still lacking.
%linked to the behaviour of the Hessian matrix $\nabla^2 L$ around the minimum, 
Nevertheless, several approaches to quantify flatness and sharpness have been developed over the years, but they have failed to paint a complete picture of the generalisation capabilities based on geometry, as a universal correlation between flatness and generalisation has been disputed \cite{zhang2021flatness, granziol2020flatness}.
In particular authors in \cite{dinh2017sharp} claim that the conclusion that flat minima should generalise better than sharp ones cannot be applied as is without further context. 
Likewise, Andriushchenko et al.\cite{andriushchenko2023modern} recently observed in multiple cases that sharper minima can generalise better in some modern experimental settings.
%the existence of a positive correlation of sharpness and generalisation implying that sharper minima can generalise better in some modern experimental settings.}

Arguably, the most impactful sharpness measure, the $\epsilon$-sharpness, was introduced by Keskar et al.\ \cite{keskar2017on}. 
It decodes the information from the eigenvalues of the Hessian matrix, while at the same time avoiding the computation-heavy calculation of the Hessian matrix itself.
Alternative measures of sharpness include
the consideration of local entropy around a minimum \cite{chaudhari2019entropy} or of the size of the connected region around the minimum where the loss is relatively similar \cite{hochreiter1997flat}.
%Keskar et al., however, introduced the seemingly most impactful approach, $\epsilon$-sharpness, which decodes the information from the eigenvalues while at the same time avoiding the calculation of the Hessian \cite{keskar2017on}.
%The seemingly most impactful approach, however, is called $\epsilon$-sharpness, was introduced  by Keskar et al. \cite{keskar2017on}, and decodes the information from the eigenvalues while at the same time avoiding the calculation of the Hessian.
%This measure focuses on a small neighborhood of a minimum and computes the largest value potentially attained by the loss function. 
%Alternative measures of sharpness have been proposed: in \cite{chaudhari2019entropy} sharpness is described using local entropy around a minimum and in \cite{hochreiter1997flat}, is defined as the size of the connected region around the minimum where the loss is relatively similar.
%which on its own does not show a complete picture of the generalisation capabilities, as a strict correlation between flatness and good generlisation has been disputed \cite{zhang2021flatness, granziol2020flatness, dinh2017sharp}.
Li et al.~\cite{li2018visualizing} however show that a problem in the interpretability of sharpness measures, such as the $\epsilon$-sharpness, may lie in the scaling of the weights.
An apparent example is optimisers with weight penalties, which enforce smaller parameters, and are thus more prone to disturbance, leading to sharp minima with good generalisation.
In order to overcome this limitation, they suggest to use filter-normalisation for the visualisation of loss landscapes and argue that flatter minima in low-dimensional visualisations with filter-normalised directions go hand-in-hand with better generalisation capabilities, even when compared across different \ac{ANN} architectures.
Even though this relationship is made evident in several instances on a qualitative level, a quantitative measure of the sharpness in the context of filter-normalisation and a corresponding analysis are not provided.
%\add{Additionally, the advantage of flat minima --also considering adaptive sharpness measures-- has been disputed recently for modern architectures and in some cases, even sharp minima have shown better performance}~\cite{andriushchenko2023modern}.

Beyond, a core weakness with respect to the universal validity of the results in most previously mentioned contributions is that experiments are limited to established benchmark data sets for image classification, such as CIFAR-10 \cite{krizhevsky2009learning} or ImageNet~\cite{deng2009imagenet}, and should thus be further verified in different research areas and contexts.
In this work, we focus on exploring the \ac{ASC} task of the DCASE2020 challenge, which belongs to the same category of tasks as CIFAR-10 (10-class classification problem), but comprises a different modality (audio instead of images) and more challenges of real-world data.
The DCASE \ac{ASC} challenge has seen tremendous influence on the computer audition community \cite{Mesaros2018}. The yearly updated data sets have been the basis for \ac{ASC} studies ranging from the development of new model architectures \cite{Guo2017} and the evaluation of model robustness \cite{hu2020device, pham19b_interspeech}, to investigations of fairness in performance amongst different recording devices and locations \cite{Triantafyllopoulos2021Fairness}. 

In this contribution, we suggest a new approach to quantitatively measure the sharpness of a local minimum --or at least of the neighbourhood of a 'well-trained' model state-- and find correlations to the generalisability of \ac{ASC} models. %to unseen data in the same and new domains compared to the seen data
%thus trying to contribute to better model selection practices.
We design our experiments considering different architectures, training parameters, and optimisation algorithms in order to address the following research questions: 
% \textbf{Research Qurestions!}
\begin{itemize}
    \item Is the sharpness derived from a two-dimensional filter-normalised visualisation stable across random directions? 
    \item How does the sharpness of \ac{ASC} models correlate with the generalisation error for \ac{IID} and \ac{OOD} data?
    \item Which hyperparameters of model training are drivers for sharp minima?
    %Do second order optimisers, which take advantage of curvature for the optimisation process, impact the sharpness of the generalisation?
    %\item Does the Curvature of minima have an impact on the generalisation outside with respect to unseen microphones?
    %\item Does the Curvature of minima have an impact on the generalisation outside with respect to unseen Cities?
    % \item Can we pick a good checkpoint?
\end{itemize}
These investigations might give insights relevant to the selection of models that generalise better to \ac{OOD} data, as well as drive the understanding of different factors affecting this generalisation for computer audition, which are both important open questions for \ac{ASC}.

\section{Methodology}
\subsection{Filter-Normalisation}
The basis for our characterisation of minima are low-dimensional filter-normalised visualisations of the loss minima as introduced in \cite{li2018visualizing}. 
The prerequisite for such a visualisation is an \ac{ANN} with parameters~$\theta$, which was trained to a model state $\theta^*$, close to a local minimum of the loss function, given a training set $X$.
The precise minimum, however, will most likely not be reached in practice, given a finite time for training, finite numerical precision, and in particular, through techniques such as early stopping. 
% In most cases, this means that the loss function $L$ is converging, such that the values of the weights (or parameters) $\Theta$ of the \ac{ANN} are in a state $\theta^*$, where
% \begin{equation}
%     L(\theta^*, X) \approx 0
% \end{equation}
% with $X$ being the training data. 
The loss function around the trained model state will nevertheless in most cases increase, when varying any of the parameters $\theta_i$ of the network. With common \ac{ANN}s having millions or even billions of parameters, this leads to very high-dimensional loss landscapes. 
%In principle, positive eigenvalues of the Hessian correspond to positive curvature in the direction of the associated eigenvector. %, with slope volume analogous to the magnitude of the eigenvalue. 
%Thus, sharp minima possess a considerable number of large positive eigenvalues while flat minima go along with small ones. %However, is not clearly defined and leads to different approaches in its implementation and interpretations.
The immediate surroundings of the minimum can best be described with the Hessian matrix.
The high dimensionality however makes
%which contains 
%any
% AT:
%all second-order derivatives of the loss function with respect to the parameter-dimensions and can thus be interpreted as a matrix of curvature values. 
%However, the high dimensionality of the problem makes 
the calculation of the Hessian matrix very computation-heavy and thus not practical \cite{moosavi2019robustness}, although significant attempts are addressed in this direction \cite{yao2020pyhessian}.

\begin{figure}
     \centering
     \hspace*{\fill}
     \begin{subfigure}[b]{0.2\textwidth}
         \centering
         \includegraphics[width=\textwidth]{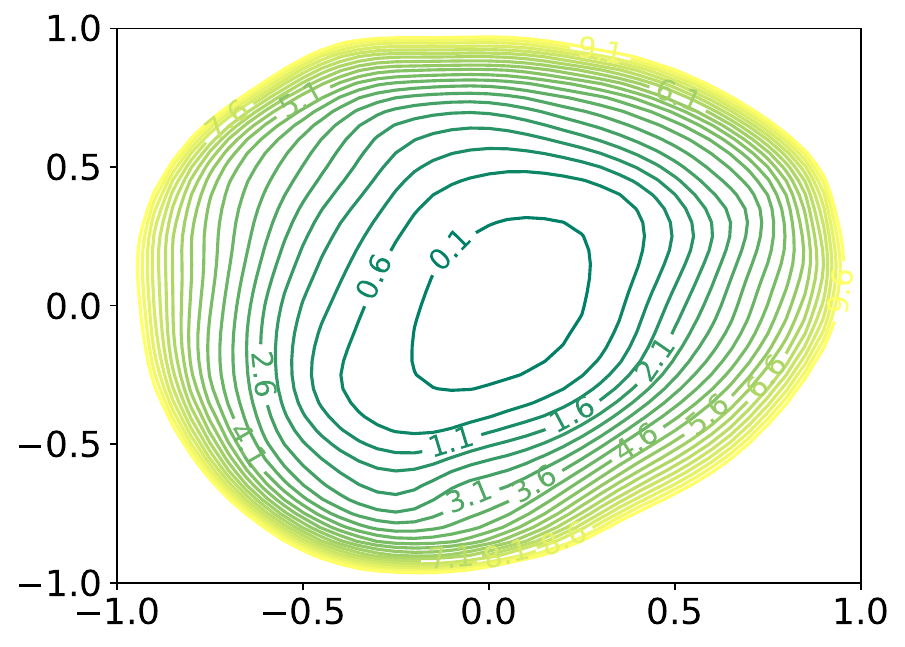}
         \caption{CNN14 with SGD optimiser and $10^{-3}$ learning rate}
         \label{fig:DCASE20_mixup-no_augment_cnn10_Adam_test}
     \end{subfigure}
     \hspace{0.4cm}
     \begin{subfigure}[b]{0.2\textwidth}
         \centering
         \includegraphics[width=\textwidth]{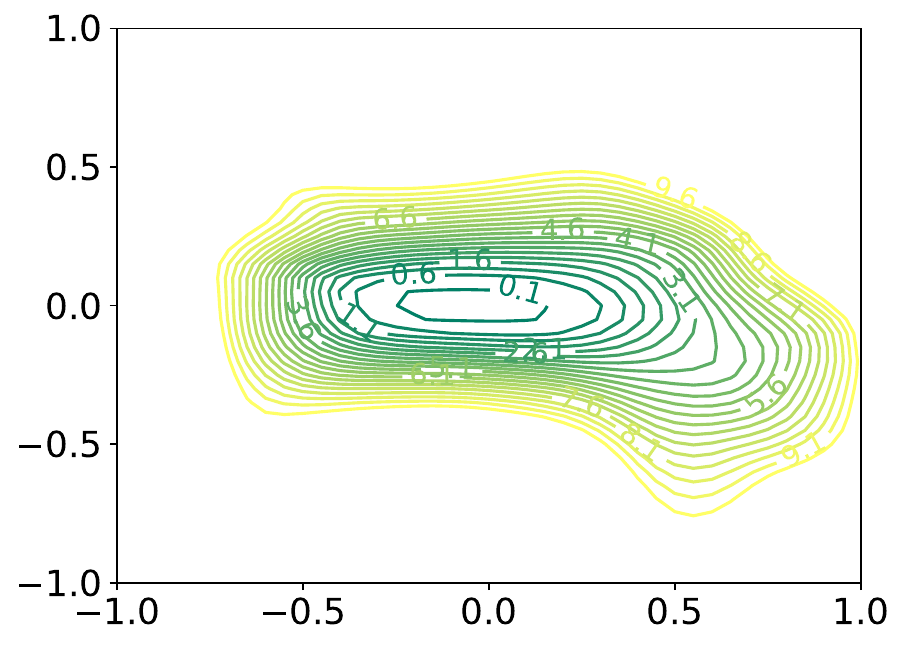}
         \caption{CNN10 with Adam optimiser and $10^{-4}$ learning rate}
         \label{fig:DCASE20_mixup-no_augment_cnn10_Adam_train}
     \end{subfigure}
     \caption{Visualisation of the two-dimensional filter-normalised loss landscape for two different model states with different architectures and training paradigms.\label{fig:visualisation}}
     %\label{fig:problem_baa500_multiple_particles}
\end{figure}

Instead, a common approach to look at the loss landscape is through low-dimensional visualisations. In two dimensions, this can be realised through the choice of random Gaussian vectors $\delta$ and $\eta$, both of the same dimension as $\theta$, which are in the following used to project the loss function as
\begin{equation}
\label{eq:visualisation}
    f(\alpha,\beta) = L(\theta^* + \alpha \delta  + \beta \eta).
\end{equation}
By varying the scalar variables $\alpha$ and $\beta$, we can depict a 2-dimensional projection of the loss landscape. 
However, Li et al.\ point out some weaknesses of the visualisation, as different models --and even different model states of the same architecture-- can have differently scaled parameters, thus making them more or less vulnerable to perturbations of the same magnitude \cite{li2018visualizing}. 
Therefore, they suggest adjusting the perturbations relative to the magnitude of the weights, thus rescaling the random gaussian directions $\delta$ and $\eta$ choosing a filter-level normalisation. 
This can be formulated as
\begin{equation}
\label{eq:filter-normalisation}
\delta_{i,j} \leftarrow \frac{\delta_{i,j}}{||\delta_{i,j}||} ||\theta_{i,j}||,
\end{equation}
where the indices of $\delta_{i,j}$ and $\theta_{i,j}$ refer to the components of $\delta$ corresponding to the $j$th filter of the $i$th layer in a convolutional neural network. \Cref{fig:visualisation} shows two examples of filter-normalised loss landscapes in 2D around a minimum with $\alpha$ and $\beta$ ranging from -1 to 1, thus varying the filters of the network by around $\pm 100\%$. 
We will use plots of this kind for the following analyses with the adapted code provided by the authors in \cite{li2018visualizing}. %\footnote{https://github.com/tomgoldstein/loss-landscape}.
As the filter-normalised plots are solving the problem of different scales of filters, the authors claim that flatter minima in this representation, despite the heavy reduction in dimensionality, indicate better generalisation, which is underlined with a qualitative analysis of several model states, trained on the CIFAR-10 dataset.

\subsection{Sharpness}
In order to quantitatively evaluate these claims for our ASC problem, we base our analysis on the $\epsilon$-sharpness, which is prominently used in the literature. 
This measure focuses on a small neighbourhood of a minimum and computes the largest value potentially attained by the loss function and is considered a good approximation of the curvature of the minimum and thus, of the sharpness or flatness of the minimum. Formally, it is defined as
%This measure focuses on the maximum loss around the minimum and is defined as 
\begin{equation}
\label{eq:sharpness}
    s_\epsilon = \frac{\mathrm{max}_{\theta \in B(\epsilon,\theta^*)}(L(\theta) - L(\theta^*))}{1+L(\theta^*)}\times 100,
\end{equation}
where \change{ $\epsilon$ determines the radius of the ball $B(\epsilon,\theta^*)$ around $\theta^*$}{$B(\epsilon,\theta^*)$ is a Euclidean ball centred on a minimum $\theta^*$ with radius $\epsilon$, i.e., \{ $\theta \in \mathbb{R}^n : \| \theta - \theta^* \| < \epsilon$ \} }. %Using the Taylor expansion and the properties of the Hessian, it can be proven that the above expression is bounded by $|\lambda_{max}|\epsilon^2,$ where $\lambda_{max}$ is the largest eigenvalue \cite{dinh2017sharp}. 
%The $\epsilon$-sharpness is thus a good approximation of the curvature of the minimum and thus, of sharpness or flatness of the minimum.

\remove{Inspired by the $\epsilon$-sharpness, we calculate the sharpness for our}\add{We follow} \eqref{eq:sharpness} \add{to calculate a quantitative sharpness measure of the} two-dimensional visualisation \add{(obtained from} \eqref{eq:visualisation} \add{and} \eqref{eq:filter-normalisation}).
%\change{based on the}{In other words, we calculate the sharpness according to} \eqref{eq:sharpness} \add{based on the} largest value \remove{of} $L(\theta)$\change{ a maximum distance of $\epsilon$ to the minimum \add{$L(\theta^*)$} of the visualisation}{ and the lowest value $L(\theta^*)$ in a 2D visualisation with a circular base}.
We will utilise this sharpness measure in the following to analyse the influences certain experimental settings have on the sharpness of minima and, further, what sharpness can tell us about the generalisation of an ASC model on unseen data. 
%, which in most cases is equal to a converging loss with value close to zero. The \ac{ANN} at this point is characterised by the values of its weights or parameters $\Theta^*$, such that  

\section{Experiments and Discussion}
%\textcolor{red}{Andreas: Please add a small description on the dataset and the Preprocessing (Mel-Spectrograms, etc.?) }
\subsection{Dataset}
As our dataset, we use the development partition of the DCASE 2020 Acoustic Scene Classification dataset~\cite{Heittola2020} and evaluate the experiments based on the standard metric accuracy, which is defined as the ratio of correctly classified samples over all samples.
The dataset includes 64 hours of audio segments from 10 different acoustic scenes, recorded in 10 European cities with 3 real devices (denoted as \textit{A}, \textit{B}, \textit{C}), as well as data from 6 simulated devices (denoted as \textit{S1}-\textit{S6}).
%Data from devices A, B, C, S1, S2, S3 are available in both training and test sets. Audio segments coming from devices S4, S5, and S6 are used only for testing.
We use the official training/evaluation splits with devices S4-S6 only appearing in the test set (\ac{OOD}).
The data is evenly distributed across cities, whereas device A (Soundman OKM II Klassik/studio A3) is dominating over B, C, and the simulated devices.
We extract 64-bin log-Mel spectrograms with a hop size of 10\,ms and a window size of 32\,ms, additionally resampling the 10\,s long audio segments to 16\,kHz.

\subsection{Model training}
Our initial experiments involved two \ac{CNN}-based architectures, the \acp{PANN} CNN10 and CNN14 \cite{kong2020panns} \add{both} with random initialisation and around 5.2 million and 80.8 million parameters, respectively, which have frequently been applied to computer audition tasks, including the DCASE \ac{ASC} task~\cite{Triantafyllopoulos2021Fairness, kong2020panns}. %, khandelwal2022fmsg}.%, mei2021encoder}.
%\textcolor{red}{Exchange reference above!!!!!!!!!!!!!!!!!!!!!!!!!!!!!!!!!!!!!!!!!!!!!!!!!!!!!!!!!}
Their convolutional nature is well in line with the CNNs for which the filter-normalisation was developed.
%Note that the focus of this work is not to produce or reproduce state-of-the-art results, but rather to give insights into common training paradigms for \ac{ASC}.
We explored widely-used optimisers, such as \ac{Adam} and \ac{SGD} with momentum, as well as less common optimisation algorithms, such as the second-order \ac{KFAC} \cite{martens2015optimizing} and \ac{GDTUO} \cite{chandra2022gradient}.
\ac{KFAC} utilises approximations to the Hessian matrix to improve convergence speed, while \ac{GDTUO} automatically adjusts hyperparameters using a stack of multiple optimisers, which in this case involves two stacked Adam optimisers, called hyperoptimisers. 
However, both KFAC and GDTUO resulted in higher computational costs in terms of runtime and memory requirements per optimisation step.
%We explore different optimisers for the models, including the very common optimisers such as Adam and \ac{SGD} optimiser with momentum, as well as the less common second-order \ac{KFAC} and the recent \ac{GDTUO}, in this case stacking two Adam optimisers on top of each other.
%The latter use approximations to the Hessian for a quicker convergence with respect to optimisation steps, even though paying with more computational costs (runtime and memory requirements) per optimisation step.
We ran a grid-search for hyperparameters as manifested in \Cref{tab:architecture},
\remove{We additionally applied a learning rate of $10^{-5}$ for the KFAC optimiser and excluded the learning rate $10^{-4}$ for CNN14 with the SGD optimiser in order to prevent suboptimal convergence. 
Given some hardware limitations for the experiments, we only utilised the second-order optimisers for the CNN10 architecture,} leading to overall 38 trained model states.
Besides the learning rate, we used default parameters for the optimisers, with SGD using a momentum of 0.9.

%We applied a grid-search for hyperparameters as manifested in \Cref{tab:architecture}.
In all cases, the training was stopped after 50 epochs and the best model state of the epoch with the highest accuracy on the development set used for testing.
%However, 
%Further, as different optimisers have different preferences on learning rates and in order to avoid the models prior to convergence, we removed the learning rate 1e-4 for CNN14 with the SGD optimiser and we added the learning 1e-5 for the \ac{KFAC} optimiser. This leads to overall 38 trained model states.
%We pre-selected trained models based on loss convergence, development set performance and relevancy for our research questions and analysed overall 24 trained model states. 
The training is implemented in \textsc{PyTorch} 1.13.1+cu117 and models were trained on a NVIDIA GeForce GTX TITAN X and a NVIDIA TITAN X (Pascal), both with 12GB RAM. 
The training time per epoch mostly varied depending on the chosen optimiser, ranging from approximately four minutes for the SGD and Adam optimisers to slightly over six minutes for KFAC, and up to around 18 minutes for GDTUO.
Our code and trained model states are publicly available\footnote{\url{https://github.com/EIHW/ASC_Sharpness}}.
%The training time per epoch depended mostly on the chosen optimiser and ranged from around four minutes (SGD, Adam), via slightly more than 6 minutes (KFAC), up to around 18 minutes (GDTUO). \textcolor{red}{Do we make the code available/did you already publish this code somewhere else, Andreas?}

\begin{table}[ht]
  \caption{Overview of the grid search parameters for model training.}
  \label{tab:architecture}
  \centering
  \begin{tabular}{lc}
    %\toprule
    % Parameter  & Values\\
    %\midrule
     Network & CNN10, CNN14\\
     Optimiser & SGD, Adam, GDTUO\footnote{Learning rate refers to the highest optimiser on the stack for GDTUO, since this is not a hyperoptimiser.}, KFAC\add{\footnote{GDTUO and KFAC are only applied to the CNN10 architecture, due to hardware limitations.}}\\
     Learning Rate & $10^-3$, $10^-4$\add{\footnote{Not applied to SGD due to suboptimal convergence.}}, $10^-5$\add{\footnote{Only applied to KFAC due to suboptimal convergence of other optimisers.}}\\
     Batch Size & 16, 32\add{\footnote{Not applied to KFAC due to hardware limitations.}}\\
     Random Seeds & 42, 43
  \end{tabular}
\end{table}

\subsection{On the robustness towards random directions}
\label{sec:robustness}
Even though not emphasised by the authors of the filter-normalisation method, the choice of the random Gaussian direction should have some impact on the measured or perceived sharpness of a given minimum. To mitigate this impacts in similar settings the authors in \cite{horoi2022exploring} use more directions in the parameter space, while in \cite{bottcher2022visualizing}, it is suggested to analyse projections along Hessian directions as an alternative method.
%Nevertheless, sharp should show a sharp behaviour in most random directions, as the trajectory is generally affected by all eigenvalues of the Hessian, thus allowing for insights gained from low-dimensional representations.
Nevertheless, most interpretations of the sharpness of minima are limited to (statistics of) a low-dimensional analysis and  often show  consistent trends across different random directions \cite{goodfellow2014qualitatively}, \cite{wu2020adversarial}, \cite{smith2017exploring}. 
%should show a sharp behaviour in most random directions, as the trajectory is generally affected by all eigenvalues of the Hessian, thus allowing for insights gained from low-dimensional representations.
We tested the robustness of our sharpness measure by calculating it based on three plots with different random directions.
In order to stay in line with the visual argumentation of the plots, as well as the characteristics of the filter-normalisation, we chose a neighbourhood of radius $0.25$ to calculate the sharpness.
Due to the high computational costs of such visualisations, the resolution was set to 0.025 in each direction, leading to 121 loss values per visualisation.
The time required to compute one sharpness value in this scenario is around 45 minutes
on a single NVIDIA A40 GPU with 16GB RAM.

\begin{figure}
     \centering
     % \hspace*{\fill}     
     \centering     \includegraphics[width=6.2cm,keepaspectratio]{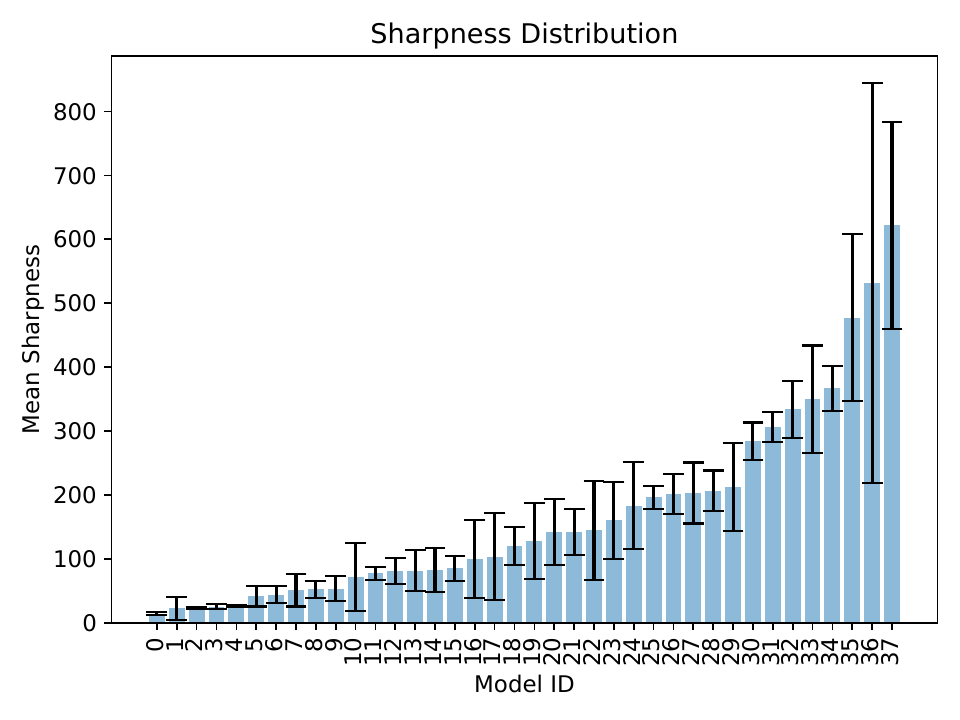}
     \caption{Distribution of sharpness-measures. Each bar indicates the mean sharpness value with the standard deviation of a trained model state in three two-dimensional plots with different random directions.\label{fig:histogram}}
     %\label{fig:problem_baa500_multiple_particles}
\end{figure}

\Cref{fig:histogram} shows the mean sharpness and standard deviation for each trained model based on three different plots per model.
Most model states show a relatively low standard deviation compared to the mean sharpness, allowing us to further interpret the sharpness in different settings.
A few exceptions with high standard deviations indicate some limitations of this approach, which might, however, be mitigated by sampling more sharpness-measures per model.
%there is some noteworthy variation between the different random directions. 
%Overall the models still seem separable enough with respect to their sharpness, allowing us to further interpret the sharpness in different settings, even though some caution needs to make. 
Similar analyses of the stability of sharpness-measures with respect to different random directions have previously been reported  \cite{wu2020adversarial}. 

\subsection{On the impact of sharpness on generalisation}
\begin{figure}
     \centering
     %\hspace*{\fill}     
     \centering
     \includegraphics[width=\columnwidth,keepaspectratio]{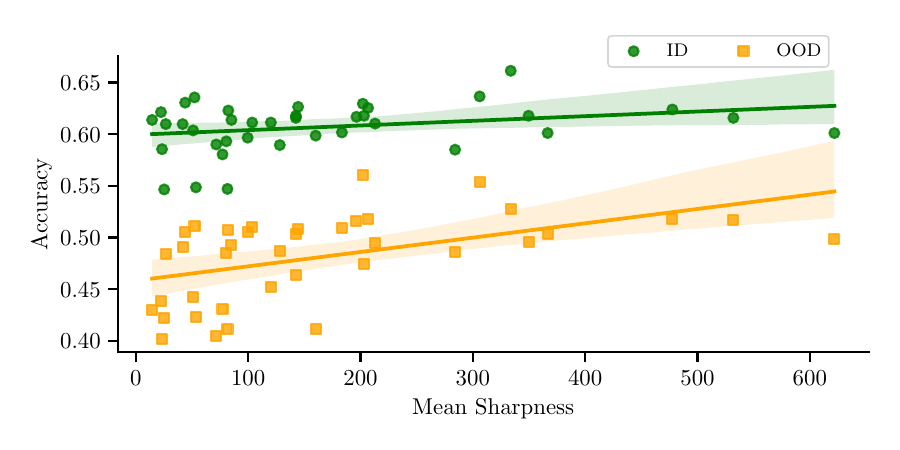}
     \caption{
     Correlation plot between sharpness of minima (the higher, the sharper) and test accuracy for all trained models.
     Showing best-fit line and 95\% confidence intervals for different models.
     \label{fig:linreg_regular}}
     %\label{fig:problem_baa500_multiple_particles}
\end{figure}

In order to gain insights into the generalisation capabilities of flat and sharp minima in \ac{ASC}, we plot the test accuracies of the trained model states against their mean sharpness value in \Cref{fig:linreg_regular}.
We thereby consider the accuracy for \ac{IID} and \ac{OOD} separately. 
To that end, we define \ac{OOD} performance as the accuracy evaluated on the devices not represented in the training data, namely S4, S5, and S6, whilst \ac{IID} performance is evaluated on the devices A, B, C, S1, S2 and S3, which are known at training time.
%In order to gain insights into the generalisation capabilities of flat and sharp minima in \ac{ASC}, we plot the test accuracies of the trained model states against their mean and maximum sharpness value in \Cref{fig:linreg_regular}.
Note that all discussed model states show a nearly 100\% accuracy on the training data, such that one minus the test accuracy can be interpreted as the generalisation gap.
Firstly, we note a tendency that, in our experiments, sharper minima show a better generalisation than flat minima.
This is a rather surprising finding, as most of the existing literature reports preferable characteristics of flat minima in the computer vision domain, e.g., \cite{hochreiter1997flat}, \cite{chaudhari2019entropy}, \cite{keskar2017on}, \cite{izmailov2018averaging}, \cite{he2019asymmetric}, \cite{zhou2020towards},
\cite{stutz2021relating}, whilst only few studies report on good generalisation in context of sharp minima \cite{tartaglione2020pruning, andriushchenko2023modern}. % with a \ac{PCC} of 0.41 between the mean sharpness and the test error, and 0.38 for the maximum sharpness, respectively.
Further investigations are necessary to unravel, whether %a general difference in the impact of sharpness on generalisation can be found for computer audition models compared to computer vision models. 
our results are an indication of a general disparity of the impact of sharpness on generalisation in acoustic scene classification and image classification. 
Critical differences in the learning of computer audition models compared to computer vision models have been reported in our previous work%, as for the former, when fine-tuning a CNN, the first layers 
: when fine-tuning a CNN for a computer audition task, the first layers were subject to more changes than the later layers \cite{triantafyllopoulos2021role}. This finding contradicts the common understanding, resulting from computer vision analyses, of earlier filters being trained to recognise low-complexity objects, such as edges, and are thus transferable without major changes amongst different tasks. 
 %\remove{Nevertheless, as it is pointed out in 
 %\cite{dinh2017sharp}
 %, certain aspects of high-dimensional geometries can change the view on sharpness. This was in particular discussed for networks with rectified linear units, such as CNN10 and CNN14, which foster that many parameter states produce the same predictive observations.
%Sharper minima could thus lie in areas, that cover a larger range of observational diversity, and thus of approximation capabilities.}
  %a more clear analysis is necessary.

%Moreover, we investigated the impact of sharpness on \ac{OOD} generalisation.
Moreover, this effect seems to be considerably higher for \ac{OOD} accuracy compared to \ac{IID} accuracy, as we observe a correlation of $.49$ in the former and a correlation of $.28$ in the latter case. 
%\cref{fig:linreg_regular} shows  over mean sharpness, with the two values showing a moderate correlation of $.49$.
%This indicates that sharper minima yield better \ac{OOD} generalisation; in fact, this correlation is considerably higher than the in-domain one of $.28$.
Based on our exploratory analysis, we hypothesise that flatter minima are over-optimised for the \ac{IID} devices --in particular, device A which dominates the training set-- and thus fail to generalise well to unseen devices.
Nevertheless, the reasons for positive correlations between sharpness and generalisation are not obvious at this moment and should be further looked into.

% Observations we need to/can address
% \begin{itemize}
% \item It seems that here sharper minima generalise better, even though very conclusive result. $\rightarrow$ further firing up the discussion on flat vs. sharp minima.
% \item Robustness towards random directions is there to a certain degree
% \item Different impact on sharpness by different hyperparameters.
% \item How well does it work with unseen devices
% \end{itemize}

%\subsection{On the impact of shaprness on OOD Generalisation}
%Moreover, we investigated the impact of sharpness on \ac{OOD} generalisation.
%To that end, we define \ac{OOD} performance as the accuracy evaluated on the devices not represented in the training data, namely S4, S5, and S6.
%\cref{fig:linreg_regular} shows \ac{OOD} accuracy over mean sharpness, with the two values showing a moderate correlation of $.49$.
%This indicates that sharper minima yield better \ac{OOD} generalisation; in fact, this correlation is higher than the in-domain one of $.28$.
%We hypothesise that flatter minima are over-optimised for the in-domain devices -- in particular, device A which dominates the training set -- and thus fail to generalise well to unseen devices.

\subsection{On the impact of hyperparameters on sharpness}
\begin{figure}
     \centering
     \begin{subfigure}[b]{0.4\textwidth}
         %\hspace{-0.5cm}
         %\centering
         \includegraphics[width=6cm,keepaspectratio]{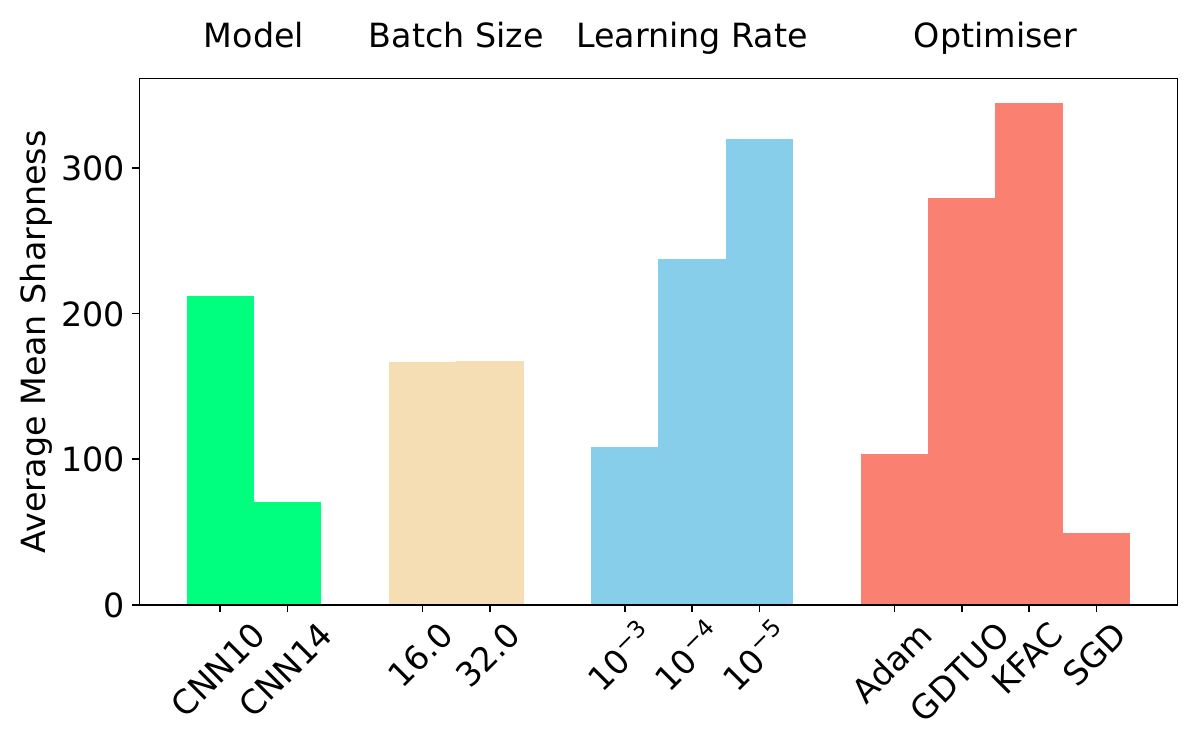}
         %\caption{}
         \label{fig:Disaggregated_Sharpness}
     \end{subfigure}
     % \hspace{0.4cm}
     \begin{subfigure}[b]{0.4\textwidth}
     %\hspace{-0.5cm}
         %\centering
         \includegraphics[width=6cm,keepaspectratio]{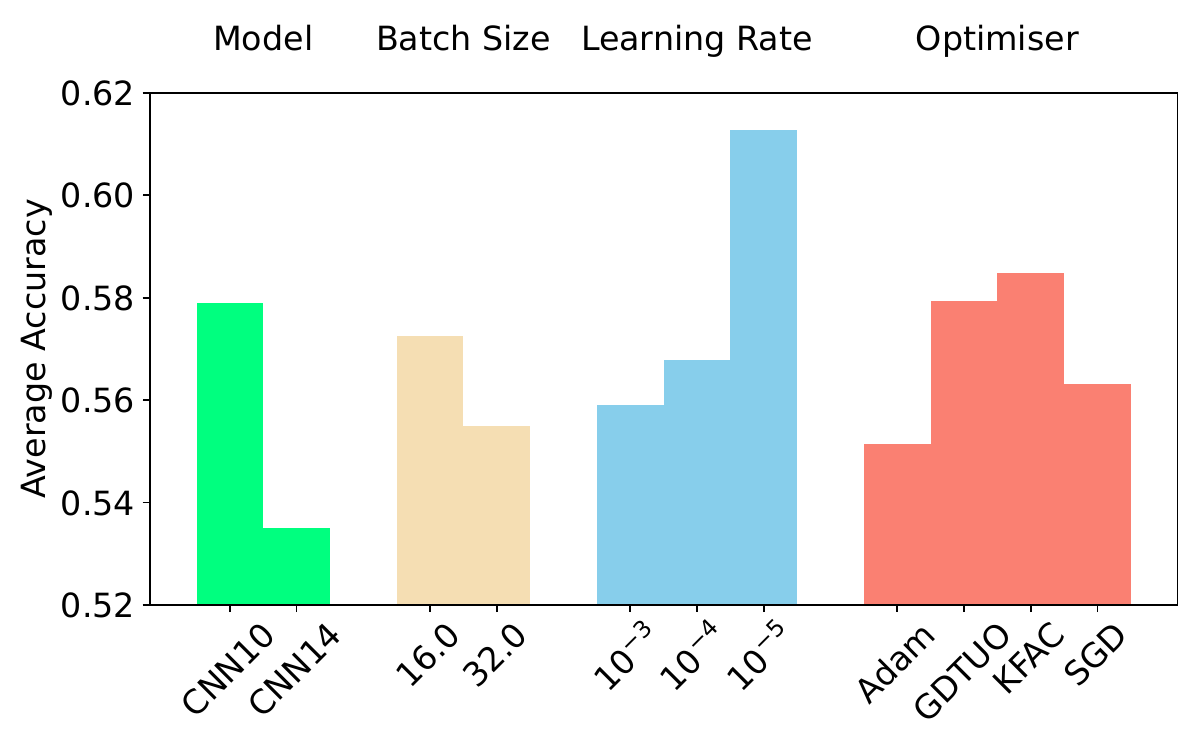}
         %\caption{}
         \label{fig:Disaggregated_Accuracies}
     \end{subfigure}
     \caption{Disaggregated distribution of mean sharpness and accuracy across hyperparameters. Each bar averages the mean sharpness or accuracy of all trained models states, grouped by the different types of hyperparameters.\label{fig:histogram_disaggregated}}
     %\label{fig:problem_baa500_multiple_particles}
\end{figure}

As a final aspect, we analyse the impact of the choice of different hyperparameters or experimental settings on the sharpness and compare these to the corresponding impact on test accuracy. \Cref{fig:histogram_disaggregated} 
suggests that both sharpness and accuracy are similarly affected by the training parameters. 
Certain hyperparameters lead to a higher value in both subplots compared to the other hyperparameters in the group, except for the batch size.
%suggest that both, the sharpness, as well as the accuracy are similarly affected by the training parameters, with a certain hyperparameter leading to a higher bar in both subplots when compared to the other hyperparameters in the group (except for the batch size).%
This result is in line with our previous findings of sharper minima tending to have better generalisation.
However, upon closer examination, it becomes apparent that the amount by which both subplots are affected by a certain group can vary considerably, as the selection of optimisers seems to have the highest impact on sharpness, which is not the case for the test accuracy.
This provides us with some insights about when a deduction of generalisation from sharpness might be more reasonable, as, for instance, different optimisers seem to bring different tendencies in sharpness, which might not fully translate to generalisation.
A remarkable similarity between average mean sharpness and average test accuracy can, however, be observed for the two model architectures, whose sharpness derives from a different(-dimensional) loss landscape. 
Note that the choice of learning rates and optimisers were not independent of each other, which limits their separate expressiveness.
%, some caution is necessary when trying to deduce the capabilities for generalisation from the sharpness of minima, especially when  
%

% \begin{figure}
%      \centering
%      \hspace*{\fill}     
%      \centering
%      \includegraphics[width=8cm,keepaspectratio]{figs/histogram_criterion.pdf}
%      \caption{Distribution of sharpness across hyperparameters. Each bar averages the mean sharpness of all trained models states, grouped by the different types of hyperparameters.\label{fig:histogram}}
%      %\label{fig:problem_baa500_multiple_particles}
% \end{figure}

%\subsection{On the impact of generalisation and fairness}
%\vspace{-1em}
\subsection{Limitations}
% \noindent
% \textbf{Limitations: } 
One of the limitations of our approach lies in the robustness of the sharpness measure, which might, however, be overcome by more efficient implementations, allowing for the consideration of additional random directions.
Beyond that, a more thorough analysis of the convergence status of models and its impact on the sharpness measure and generalisation seems desirable.
Especially, considering that not all experimental details could be investigated in depth, this contribution can only be a piece in the debate about flat versus sharp minima in ASC in particular and computer audition in general.
Beyond, the reasons for good generalisation capabilities of sharp minima in our exploratory study need to be further investigated as the impact of individual hyperparameters on the training needs to be better understood.
%on this and other real-world datasets
%Overall, our approach shows certain limitations. 
%The first, being already extensively discussed in section in \Cref{sec:robustness}, is the robustness towards random directions.
%Beyond, the rather low resolution of the visualisation plot--chosen due to computational costs-- might have an impact on the sharpness measure.
%Future approaches offering computationally practical consideration of more dimensions or a higher resolution might be preferable in that aspect, producing an even more reliable sharpness measure.

%We further highlight that we evaluate models under practical training conditions. This means in particular that we often choose model states before the final epoch to avoid overfitting. 
%Thus, most models were likely only close to the local minimum they would potentially converge to, which might have a certain effect on the calculation and interpretation of the sharpness.
%For this reason, we also excluded some model states, which were not close enough to convergence. 
%A more refined approach to select non-converged models or alternatively, to consider the (expected) distance to a local minimum seems desirable.

%Beyond, even though we carried out extensive experimentation on the given task, only a subset of all relevant aspects could be investigated in depth.
%This contribution can therefore only be a piece in the debate about flat versus sharp minima in ASC.

%\vspace{-1em}
\section{Conclusions} 
In this contribution, we explored the sharpness of minima in the loss function for acoustic scene classification models and its impact on the generalisation capabilities in different, practice-relevant, experimental settings.
We found that for our trained models, sharper minima generalised better to unseen (in particular to OOD) data, which has rarely been observed in the computer vision domain.
Our approach shows some limitations, as for instance, the choice of optimisers has a higher impact on the sharpness of minima than on the generalisation.
In future work, we plan to focus on more efficient and interpretable implementations of sharpness measures and to better understand individual effects of hyperparameters before our findings can be put into practice.

\section{Acknowledgements}
This work was partially funded by the DFG's Reinhart Koselleck project No.\ 442218748 (AUDI0NOMOUS).

%BS: Commented out for now, as eats up space - take in again for camera :) please
%TODO: Open points from submission, not addressed in the paper yet:
% Hyperparameters best performing model

\bibliographystyle{IEEEbib}
%\bibliography{strings,refs}
% \bibliography{mybib}
% TODO: Changes here. Check with Andreas if alright!
{\footnotesize
\bibliography{mybib}}

\end{document}